\documentclass[aps,prb,twocolumn,showpacs]{revtex4}

\usepackage{graphicx}
\usepackage{amsmath}
\usepackage{amssymb}
\usepackage{appendix}

\begin{document}

\title{Doping dependence of Meissner effect in triangular-lattice superconductors}

\author{Ling Qin, Xixiao Ma, and L\"ulin Kuang}

\affiliation{Department of Physics, Beijing Normal University, Beijing 100875, China}

\author{Jihong Qin}

\affiliation{Department of Physics, University of Science and Technology Beijing, Beijing 100083, China}

\author{Shiping Feng$^{*}$}

\affiliation{Department of Physics, Beijing Normal University, Beijing 100875, China ~~}


\begin{abstract}
In the spin-excitation-mediated pairing mechanism for superconductivity, the geometric frustration effects not only the spin configuration but also the superconducting-state properties. Within the framework of the kinetic-energy-driven superconducting mechanism, the doping and temperature dependence of the Meissner effect in triangular-lattice superconductors is investigated. It is shown that the magnetic-field-penetration depth exhibits an exponential temperature dependence due to the absence of the d-wave gap nodes at the Fermi surface. However, in analogy to the dome-like shape of the doping dependence of the superconducting transition temperature, the superfluid density increases with increasing doping in the lower doped regime, and reaches a maximum around the critical doping, then decreases in the higher doped regime.
\end{abstract}



\maketitle

\section{Introduction}

Investigation of oxide compounds has uncovered many unusual properties characterized by the strong electron correlation, which include unconventional superconductivity and anomalous properties in the normal-state \cite{Kastner98,Aoki04}. Superconductivity in cuprate superconductors results from some special microscopic conditions \cite{Kastner98,Aoki04,Bednorz86}: (a) the one-half spin Cu ions situated in a square-planar arrangement and bridged by oxygen ions; (b) the weak coupling between neighboring layers; and (c) the charge-carrier doping in such a way that the Fermi level lies near the middle of the Cu-O $\sigma^{*}$ bond. One common feature of cuprate superconductors is the {\it square-planar} Cu arrangement \cite{Kastner98,Aoki04,Bednorz86}. However, some oxide materials with a two-dimensional spin arrangements on non-square lattices have been synthetized \cite{Aoki04,Ramirez94}. In particular, it has been reported \cite{Takada03,Schaak03,Milne04,Sakurai06,Michioka06} that there is a class of cobaltate superconductors Na$_{x}$CoO$_{2}\cdot y$H$_{2}$O, which have a lamellar structure consisting of the two-dimensional CoO$_{2}$ layers separated by a thick insulating layer of Na$^{+}$ ions and H$_{2}$O molecules, where the one-half spin Co$^{4+}$ ions sites sit not on a square-planar, but on a {\it triangular-planar lattice}, therefore allowing a test of the effect of the geometric frustration on superconductivity \cite{Aoki04,Ramirez94,Takada03,Schaak03,Milne04,Sakurai06,Michioka06}. Moreover, Na$_{x}$CoO$_{2}\cdot y$H$_{2}$O is viewed as an electron-doped Mott insulator, where superconductivity appears with electron doping \cite{Takada03,Schaak03,Milne04,Sakurai06,Michioka06}. Furthermore, it has been found that the antiferromagnetic (AF) spin correlation \cite{Kobayashi03,Fujimoto04,Ihara06,Kato06,Zheng06} is present in Na$_{x}$CoO$_{2}\cdot y$H$_{2}$O, although being much weaker than those in square-lattice cuprate superconductors. In this case, a question is whether the unusual features observed on square-lattice superconductors exist also in triangular-lattice superconductors or not? The finding of superconductivity in triangular-lattice cobaltate superconductors has raised the hope that it may help solve the unusual physics in square-lattice cuprate superconductors. On the other hand, the doped Mott insulator on a triangular lattice is also of interests in its own right with many unanswered fascinating questions \cite{Aoki04,Starykh15}, where the geometric frustration was expected to destroy the AF long-range order (AFLRO) and leads to a quantum spin-liquid state.

Superconductivity is characterized by exactly zero electrical resistance and expulsion of magnetic fields occurring in superconductors when cooled below $T_{\rm c}$. The later phenomenon is so-called Meissner effect \cite{Schrieffer64}, i.e., a superconductor is placed in an external magnetic field ${\rm B}$ smaller than the upper critical field ${\rm B}_{\rm c}$, the magnetic field ${\rm B}$ penetrates only to a penetration depth $\lambda$ and is excluded from the main body of the system. This magnetic-field-penetration depth is a fundamental parameter of superconductors, and provides a rather direct measurement of the superfluid density $\rho_{\rm s}$ ($\rho_{\rm s}\equiv\lambda^{-2}$) \cite{Schrieffer64}, which is proportional to the squared amplitude of the macroscopic wave function. In particular, the variation of the magnetic-field-penetration depth as a function of doping and temperature gives the information about the nature of quasiparticle excitations and their dynamics. Moreover, the magnetic-field-penetration depth can be also used as a probe of the pairing symmetry, since it can distinguish between a fully gapped and a nodal quasiparticle excitation spectrum \cite{Schrieffer64,Bonn96,Tsuei00}. The former results in the thermally activated (exponential) temperature dependence of the magnetic-field-penetration depth, whereas the latter one implies a power law behavior. For square-lattice superconductors, the Meissner effect has been studied experimentally \cite{Uemura91,Suter04} as well as theoretically \cite{Yip92,Feng10}. In particular, the electromagnetic response in square-lattice superconductors has been discussed \cite{Feng10} based on the kinetic-energy-driven superconducting (SC) mechanism \cite{Feng0306,Feng15}, and the obtained results of the doping and temperature dependence of the magnetic-field-penetration depth and superfluid density are well consistent with the experimental data observed on square-lattice superconductors \cite{Uemura91,Suter04}. In triangular-lattice cobaltate superconductors, on the other hand, although the Meissner effect has been investigated by virtue of systematic studies using the muon-spin-rotation measurement technique \cite{Kanigel04,Uemural04}, the electromagnetic response has not been clarified starting from a microscopic SC theory, and no explicit calculations of the evolution of the superfluid density with doping and temperature has been made so far. In this paper, we try to study this issue within the framework of the kinetic-energy-driven SC mechanism. We show that the magnetic-field-penetration depth of triangular-lattice superconductors exhibits an exponential temperature dependence due to the absence of the d-wave gap nodes. However, in analogy to the case of square-lattice superconductors, the superfluid density in triangular-lattice superconductors also has a dome-like shape of the doping dependence.

The rest of this paper is organized as follows. The basic formalism is presented in section \ref{framework}, where we generalize the response kernel function obtained within the framework of the kinetic-energy-driven SC mechanism from the case in the previous square-lattice superconductors \cite{Feng10} to the present case for triangular-lattice superconductors, and then employ this response kernel function to obtain explicitly the doping dependence of the Meissner effect in triangular-lattice superconductors for all the temperature $T\leq T_{c}$. Based on this theoretical framework of the electromagnetic response, we then discuss the basic behavior of triangular-lattice superconductors in a weak electromagnetic field in section \ref{meissner-effect}. Finally, we give a summary in section \ref{conclusions}.

\section{Theoretical framework}\label{framework}

In triangular-lattice cobaltate superconductors, the common feature is the presence of the CoO$_{2}$ plane \cite{Takada03,Schaak03,Milne04,Sakurai06,Michioka06}, and then it is thus believed that the nonconventional SC mechanism in triangular-lattice cobaltate superconductors and the related anomalous properties in the normal-state are dominated by this plane. In this case, many authors have argued that the essential physics of the CoO$_{2}$ plane is contained in the $t$-$J$ model on a triangular lattice  \cite{Baskaran03}. To study the electromagnetic response in triangular-lattice cobaltate superconductors, the $t$-$J$ model can be generalized by including the exponential Peierls factors as,
\begin{eqnarray}\label{tjham}
H&=&t\sum_{l\hat{\eta}\sigma}e^{-i(e/\hbar){\bf A}(l)\cdot\hat{\eta}}PC_{l\sigma}^{\dagger}C_{l+\hat{\eta}\sigma}P^{\dagger}\nonumber\\
&-&\mu\sum_{l\sigma}P C_{l\sigma}^{\dagger}C_{l\sigma} P^{\dagger}+J\sum_{l\hat{\eta}}{\bf S}_{l} \cdot{\bf S}_{l+\hat{\eta}},
\end{eqnarray}
where the electron hopping integral $t<0$, the summation is over all sites $l$, and for each $l$, over its nearest-neighbor $\hat{\eta}$, $C^{\dagger}_{l\sigma}$ and $C_{l\sigma}$ are operators that respectively create and annihilate electrons with spin $\sigma$, ${\bf S}_{l}=(S^{\rm x}_{l},S^{\rm y}_{l},S^{\rm z}_{l})$ are spin operators, $\mu$ is the chemical potential, and the projection operator $P$ removes zero occupancy in the case of the electron doping, i.e., $\sum_{\sigma}C^{\dagger}_{l\sigma}C_{l\sigma}\geq 1$, while the exponential Peierls factor account for the coupling of electron charge to the weak external magnetic field in terms of the vector potential ${\bf A}(l)$. This $t$-$J$ model (\ref{tjham}) is the strong coupling limit of the Hubbard model and the crucial difficulty of its solution lies in enforcing the local constraint of no zero electron occupancy. In the case of the hole doping, an intuitively appealing approach to implement the local constraint of no double electron occupancy and the charge-spin separation scheme is the slave-particle formalism \cite{Zou88}, however, the local constraint of no double electron occupancy is explicitly replaced by a global constraint in the actual calculations. Following the charge-spin separation scheme, a fermion-spin theory has been developed \cite{Feng15,Feng0494}, where the local constraint of no double electron occupancy can be treated properly in the actual calculations. To apply the fermion-spin theory to the case of the electron doping, the $t$-$J$ model (\ref{tjham}) can be rewritten in terms of a particle-hole transformation $C_{l\sigma}\rightarrow f^{\dagger}_{l-\sigma}$ as \cite{Liu05},
\begin{eqnarray}\label{tjham-hole}
H&=&-t\sum_{l\hat{\eta}\sigma}e^{-i(e/\hbar){\bf A}(l)\cdot\hat{\eta}}f_{l\sigma}^{\dagger}f_{l+\hat{\eta}\sigma}+\mu\sum_{l\sigma}f_{l\sigma }^{\dagger}f_{l\sigma }\nonumber\\
&+& J\sum_{l\hat{\eta}}{\bf S}_{l}\cdot{\bf S}_{l+\hat{\eta}},
\end{eqnarray}
where $f^{\dagger}_{l\sigma}$ ($f_{l\sigma}$) is the hole creation (annihilation) operator. This $t$-$J$ model (\ref{tjham-hole}) in the hole representation is subject to the local constraint that double occupancy of a site by two fermions of opposite spins is not allowed, i.e., $\sum_{\sigma}f^{\dagger}_{l\sigma}f_{l\sigma}\leq 1$. The physics of the no double electron occupancy in the fermion-spin theory is taken into account by representing the fermion operator $f_{l\sigma}$ as a composite object created by \cite{Feng15,Feng0494},
\begin{eqnarray}\label{fermion-spin}
f_{l\uparrow}=a^{\dagger}_{l\uparrow}S^{-}_{l},~~~~ f_{l\downarrow}=a^{\dagger}_{l\downarrow}S^{+}_{l},
\end{eqnarray}
where the spinful fermion operator $a_{l\sigma}=e^{-i\Phi_{l\sigma}}a_{l}$ describes the charge degree of freedom of the hole together with some effects of spin configuration rearrangements due to the presence of the doped electron itself (charge carrier), while the spin operator $S_{l}$ represents the spin degree of freedom of the hole, then the local constraint of no double occupancy is satisfied in the actual calculations. In this fermion-spin representation (\ref{fermion-spin}), the $t$-$J$ model (\ref{tjham-hole}) can be expressed as \cite{Liu05},
\begin{eqnarray}\label{cssham}
H&=&t\sum_{l\hat{\eta}}e^{-i(e/\hbar){\bf A}(l)\cdot\hat{\eta}}(a^{\dagger}_{l+\hat{\eta}\uparrow}a_{l\uparrow}S^{+}_{l}S^{-}_{l+\hat{\eta}}+a^{\dagger}_{l+\hat{\eta}\downarrow} a_{l\downarrow}S^{-}_{l}S^{+}_{l+\hat{\eta}})\nonumber\\
&-&\mu\sum_{l\sigma}a^{\dagger}_{l\sigma}a_{l\sigma}+J_{\rm{eff}}\sum_{l\hat{\eta}}{\bf{S}}_{l}\cdot {\bf{S}}_{l+\hat{\eta}},
\end{eqnarray}
where $J_{\rm eff}=(1-\delta)^{2}J$, and $\delta=\langle a^{\dagger}_{l\sigma}a_{l\sigma}\rangle=\langle a^{\dagger}_{l}a_{l}\rangle$ is the charge-carrier doping concentration. At half-filling, the $t$-$J$ model (\ref{cssham}) is reduced to the AF Heisenberg model on a triangular lattice. In the early days of the spin-liquid, it was proposed that the strong geometry frustration in the triangular-lattice Heisenberg model may completely destroy AFLRO \cite{Anderson73}. Later, a series of studies with spin-wave calculations \cite{Zhitomirsky13,Jolicoeur89} and numerical simulations \cite{Bernu94,Caprioti99} indicate that the triangular-lattice AF Heisenberg model appears to have better state with three sublattice magnetic order. However, for the case in the square-lattice, it has been shown that AFLRO is destroyed by charge-carrier doping with $\delta\sim 0.05-0.07$ for $t/J\sim 2.5 -5$ \cite{Lee88,Khaliullin93}. It is thus possible that the spin-liquid state is attained in the triangular-lattice system for sufficiently low doping, such as $\delta\sim 0.05$, due to the strong geometry frustration, and then there is no AFLRO in the doped regime where superconductivity appears.

Much of the interest in oxide superconductors is due to the fact that these materials represent novel SC mechanism for superconductivity. Based on the $t$-$J$ model in the fermion-spin representation, we have developed a kinetic-energy-driven SC mechanism \cite{Feng15,Feng0306} in the case without AFLRO for a microscopic description of the SC-state of square-lattice cuprate superconductors. This kinetic-energy-driven superconductivity is purely electronic without phonons, where the charge-carrier pairing interaction arises directly from the kinetic energy by the exchange of spin excitations in the higher powers of the doping concentration. In particular, the kinetic-energy-driven SC-state is controlled by both the SC gap and quasiparticle coherence, then the maximal SC transition temperature $T_{\rm c}$ occurs around the optimal doping, and decreases in both the underdoped and overdoped regimes. On the other hand, since the strong electron correlation is common for both these materials \cite{Kastner98,Aoki04,Bednorz86,Ramirez94,Takada03,Schaak03,Milne04,Sakurai06,Michioka06,Kobayashi03,Fujimoto04,Ihara06,Kato06,Zheng06,Starykh15}, these two oxide systems may have similar underlying SC mechanism, i.e., it is possible that superconductivity in triangular-lattice cobaltate superconductors is also driven by the kinetic energy. In this case, superconductivity in triangular-lattice superconductors has been discussed based on the kinetic-energy-driven SC mechanism \cite{Liu05}, where although the effect from the charge-carrier quasiparticle coherence has been dropped, the obtained doping dependence of $T_{\rm c}$ is qualitative agreement with experimental data \cite{Schaak03,Milne04,Sakurai06,Michioka06} of Na$_{x}$CoO$_{2}$$\cdot y$H$_{2}$O. In this paper, we generalize the formalism of the kinetic-energy-driven SC mechanism developed in Ref. \onlinecite{Liu05} by considering the effect from the charge-carrier quasiparticle coherence, and then apply this new form to study the electromagnetic response in triangular-lattice superconductors. Following the previous discussions \cite{Feng0306,Feng15,Liu05}, the self-consistent equations that are satisfied by the full charge-carrier diagonal and off-diagonal Green's functions of the triangular-lattice $t$-$J$ model (\ref{cssham}) in the SC-state at zero magnetic field can be obtained explicitly as,
\begin{subequations}\label{Green-functions}
\begin{eqnarray}
g({\bf k},\omega)&=&g^{(0)}({\bf k},\omega)+g^{(0)}({\bf k},\omega)[\Sigma^{({\rm a})}_{1}({\bf k},\omega)g({\bf k},\omega)\nonumber\\
&-&\Sigma^{({\rm a})*}_{2}({\bf k},\omega) \Gamma^{\dagger}({\bf k},\omega)], ~~~~\\
\Gamma^{\dagger}({\bf k},\omega)&=&g^{(0)}(-{\bf k},-\omega)[\Sigma^{({\rm a})}_{1}(-{\bf k},-\omega)\Gamma^{\dagger}({\bf k},\omega)\nonumber\\
&+&\Sigma^{({\rm a})}_{2}({\bf k},\omega)g({\bf k},\omega)],~~~~~
\end{eqnarray}
\end{subequations}
respectively, where $g^{(0)-1}({\bf k},\omega)=\omega-\xi_{\bf k}$ is the mean-field (MF) charge-carrier Green's function, $\xi_{\bf k}=Zt\chi \gamma_{\bf k}-\mu$ is the MF charge-carrier excitation spectrum, $Z$ is the number of the nearest-neighbor sites, $\gamma_{\bf k}=(1/Z)\sum_{\hat{\eta}}e^{i{\bf k}\cdot\hat{\eta}}=[\cos k_{x}+2\cos(k_{x}/2)\cos (\sqrt{3} k_{y}/2)]/3$, and the spin correlation function $\chi$ is defined as $\chi=\langle S^{+}_{l}S^{-}_{l+\hat{\eta}}\rangle$, while the charge-carrier self-energies have been obtained as,
\begin{widetext}
\begin{subequations}\label{self-energies}
\begin{eqnarray}
\Sigma^{({\rm a})}_{1}({\bf k},i\omega_{n})&=&(Zt)^{2}{1\over N^{2}}\sum_{{\bf p},{\bf p}'}\gamma^{2}_{{\bf p}+{\bf p}'+{\bf k}}{1\over\beta}\sum_{ip_{m}}
g({\bf p} +{\bf k},ip_{m}+i\omega_{n})\Pi({\bf p},{\bf p}',ip_{m}), \\
\Sigma^{({\rm a})}_{2}({\bf k},i\omega_{n})&=&(Zt)^{2}{1\over N^{2}}\sum_{{\bf p},{\bf p}'}\gamma^{2}_{{\bf p}+{\bf p}'+{\bf k}}{1\over \beta}\sum_{ip_{m}}
\Gamma^{\dagger}({\bf p} +{\bf k},ip_{m} +i\omega_{n})\Pi({\bf p},{\bf p}',ip_{m}),~~~~~
\end{eqnarray}
\end{subequations}
\end{widetext}
with the spin bubble,
\begin{eqnarray}\label{spin-bubble}
\Pi({\bf p},{\bf p}',ip_{m})&=&{1\over\beta}\sum_{ip_{m}'}D^{(0)}({\bf p}',ip_{m}')\nonumber\\
&\times& D^{(0)}({\bf p}'+{\bf p},ip_{m}'+ip_{m}),
\end{eqnarray}
where $D^{(0)}(l-l',t-t')=\langle\langle S^{+}_{l}(t);S^{-}_{l'}(t')\rangle\rangle$ is the MF spin Green's function, and has been evaluated as,
\begin{eqnarray}\label{MF-spin-Green-function}
D^{(0)}({\bf p},\omega)={B_{\bf p}\over 2\omega_{\bf p}}\left ( {1\over \omega-\omega_{\bf p}}-{1\over \omega+\omega_{\bf p}}\right ),
\end{eqnarray}
with the function $B_{\bf p}=\lambda [2\chi^{\rm z}(\epsilon\gamma_{\bf p}-1)+\chi(\gamma_{\bf p}-\epsilon)]$, while the MF spin excitation spectrum $\omega_{\bf p}$ is given by,
\begin{eqnarray}\label{MF-spin-spectrum}
\omega^{2}_{\bf p}&=&\lambda^{2}\left [{1\over 2}\epsilon\left (A_{1}-{1\over 3}\alpha\chi^{\rm z}-\alpha\chi\gamma_{\bf k}\right)(\epsilon-\gamma_{\bf k})\right .\nonumber\\
&+&\left . \left (A_{2}-{1\over 2Z} \alpha\epsilon\chi-\alpha\epsilon\chi^{\rm z}\gamma_{\bf k}\right )(1-\epsilon\gamma_{\bf k})\right ],\nonumber\\
\end{eqnarray}
with $A_{1}=\alpha C+(1-\alpha)/(2Z)$, $A_{2}=\alpha C^{z}+(1-\alpha)/(4Z)$, $\lambda=2ZJ_{\rm eff}$, $\epsilon=1+2t\phi/J_{\rm eff}$, the charge-carrier's particle-hole parameter $\phi=\langle a^{\dagger}_{l\sigma}a_{l+\hat{\eta}\sigma} \rangle$, and the spin correlation functions $\chi^{\rm z}=\langle S_{l}^{\rm z}S_{l+\hat{\eta}}^{\rm z}\rangle$, $C=(1/Z^{2})\sum_{\hat{\eta},\hat{\eta'}}\langle S_{l+\hat{\eta}}^{+}S_{l+\hat{\eta'}}^{-}\rangle$, and $C^{\rm z}=(1/Z^{2})\sum_{\hat{\eta},\hat{\eta'}}\langle S_{l+\hat{\eta}}^{\rm z}S_{l+\hat{\eta'}}^{\rm z}\rangle$. Since the quantum spin operators obey the Pauli algebra, it needs to apply the decoupling approximation \cite{Tyablikov67,Kondo72} to the higher order spin Green's function for obtaining the MF spin Green's function $D^{(0)}({\bf p},\omega)$. In particular, in order to satisfy the sum rule of the correlation function $\langle S^{+}_{l} S^{-}_{l}\rangle=1/2$ in the case without AFLRO, an important decoupling parameter $\alpha$ has been introduced in the decoupling approximation for the higher order spin Green's function, which can be regarded as the vertex correction, and is determined self-consistently \cite{Kondo72,Feng15}.

\begin{figure}[h!]
\includegraphics[scale=0.35]{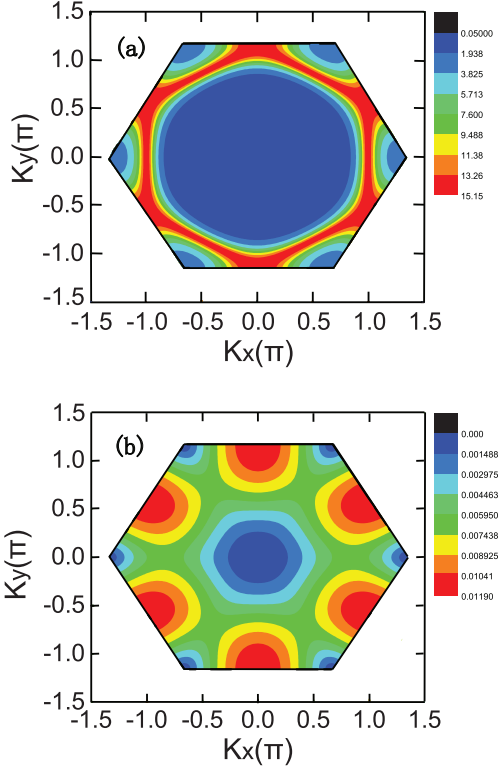}
\caption{(Color online) (a) The spectral intensity maps at the charge-carrier Fermi energy in $\delta=0.20$ with $T=0.001J$ for $t/J=-2.5$. (b) The d-wave gap maps in the Brillouin zone. \label{Fermi-surface}}
\end{figure}

In obtaining Eqs. (\ref{Green-functions}) and (\ref{self-energies}), the facts $\Sigma^{({\rm a})}_{2}(-{\bf k},-\omega)=\Sigma^{({\rm a})}_{2}({\bf k},\omega)$ and $\Gamma^{\dagger}(-{\bf k},-\omega) = \Gamma^{\dagger}({\bf k},\omega)$ have been used, which indicates that the charge-carrier pair gap $\bar{\Delta}^{({\rm a})}_{\bf k}(\omega)= \Sigma^{({\rm a})}_{2}({\bf k},\omega)$ is an even function of $\omega$. However, the other charge-carrier self-energy $\Sigma^{({\rm a})}_{1}({\bf k},\omega)$ is not. It is convenient to break it into its symmetric and antisymmetric parts as $\Sigma^{({\rm a})}_{1}({\bf k},\omega)=\Sigma^{({\rm a})}_{\rm 1e}({\bf k},\omega)+\omega\Sigma^{({\rm a})}_{\rm 1o}({\bf k},\omega)$, and then $\Sigma^{({\rm a})}_{\rm 1e}({\bf k},\omega)$ and $\Sigma^{({\rm a})}_{\rm 1o}({\bf k},\omega)$ are both even function of $\omega$. The antisymmetric part $\Sigma^{({\rm a})}_{\rm 1o}({\bf k},\omega)$ of the self-energy $\Sigma^{({\rm a})}_{1}({\bf k},\omega)$ renormalizes the MF charge-carrier spectrum, and is directly related to the charge-carrier quasiparticle coherent weight as $Z^{-1}_{\rm aF}({\bf k},\omega)=1-{\rm Re}\Sigma^{({\rm a})}_{\rm 1o}({\bf k},\omega)$, while the symmetric part $\Sigma^{({\rm a})}_{\rm 1e}({\bf k},\omega)$ of the self-energy $\Sigma^{({\rm a})}_{1}({\bf k},\omega)$ just renormalizes the chemical potential. In this paper, we mainly focus on the low-energy behavior, and in this case, the charge-carrier pair gap and quasiparticle coherent weight can be generally discussed in the static limit, i.e., $\bar{\Delta}^{(\rm a)}_{\bf k}= \Sigma^{({\rm a})}_{2}({\bf k},\omega)\mid_{\omega=0}$, and $Z^{-1}_{\rm aF}({\bf k})=1-{\rm Re}\Sigma^{({\rm a})}_{\rm 1o}({\bf k},\omega)\mid_{\omega=0}$. Although $Z_{\rm aF}({\bf k})$ still is a function of momentum, however, as a qualitative discussion, the wave vector ${\bf k}$ in $Z_{\rm aF}({\bf k})$ can be chosen in the high symmetry point of the Brillouin zone, i.e.,
\begin{eqnarray}\label{coherent-weight}
{1\over Z_{\rm aF}}=1-{\rm Re}\Sigma^{(\rm a)}_{\rm 1o}({\bf k},\omega=0)\mid_{{\bf k}_{0}},
\end{eqnarray}
with ${\bf k}_{0}=[\pi/3,\sqrt{3}\pi/3]$. In triangular-lattice superconductors, a central issue is whether the charge-carrier pair gap has nodes at the Fermi surface or not. Experimentally, it is far from reaching a consensus on the pairing symmetry. From early NMR and NQR measurements, the contradictory results were obtained, since some experimental data are consistent with the case of the existence of a coherence peak indicating a complete gap over the Fermi surface \cite{Kobayashi03,Michioka06}, while other experimental results suggest no coherence peak \cite{Fujimoto04,Zheng06}. In particular, although the recent experimental results \cite{Oeschler08} obtained from the specific-heat measurements do not give unambiguous evidence for either the presence or absence of the nodes in the energy gap, the experimental data of the specific-heat \cite{Oeschler08} are consistent with these fitted results obtained from phenomenological Bardeen-Cooper-Schrieffer (BCS) formalism with the d-wave ($d_{1}+id_{2}$ pairing) symmetry without gap nodes. On the theoretical hand, according to the irreducible representations of the triangular-lattice system, it has been pointed out that there are three possible basis functions of even parity \cite{Lee90}, i.e., one s-like function $s_{\bf k}= {\cos} k_{x}+ {\cos}[(k_{x}- \sqrt{3} k_{y})/2]+{\cos}[(k_{x}+\sqrt{3}k_{y})/2]$, and two d-like functions, $d_{1{\bf k}}=2{\cos}k_{x}-{\cos}[(k_{x}- \sqrt{3}k_{y})/2]- {\cos}[(k_{x}+\sqrt{3} k_{y}) /2]$ and $d_{2{\bf k}}=\sqrt{3}{\cos}[(k_{x}+\sqrt{3}k_{y})/2]- \sqrt{3} {\cos}[(k_{x}-\sqrt{3}k_{y})/2]$. However, with the different linear combinations of these basis functions, one find \cite{Lee90} based on the variational Monte Carlo simulation of the resonating-valence-bond wave function that the lowest energy state of the AF triangular-lattice Heisenberg model is the d-wave $(d_{1{\bf k}}+ id_{2{\bf k}})$ state with the energy gap $\Delta_{\bf k}\propto \Delta(d_{1{\bf k}}+ id_{2{\bf k}})$. Furthermore, it has been shown based on the numerical simulations that this d-wave state also is the lowest state around the electron-doped regime where superconductivity appears in triangular-lattice superconductors \cite{Honerkamp03,Watanabe04,Weber06}. In particular, the recent theoretical studies based on a phenomenological analysis \cite{Zhou08} and a combined cluster calculation and renormalization group approach \cite{Kiesel13} show that this d-wave state naturally explains some SC-state properties as indicated by experiments. In this case, we only consider the case with the d-wave pairing symmetry,
\begin{eqnarray}\label{d-wave-gap}
\bar{\Delta}^{({\rm a})}_{\bf k}=\bar{\Delta}^{({\rm a})}(d_{1{\bf k}}+id_{2{\bf k}}),
\end{eqnarray}
and then the full charge-carrier diagonal and off-diagonal Green's functions in Eq. (\ref{Green-functions}) can be obtained explicitly as,
\begin{subequations}\label{BCSGF}
\begin{eqnarray}
g({\bf k},\omega)&=&Z_{\rm aF}\left ({U^{2}_{{\rm a}{\bf k}}\over\omega-E_{{\rm a}{\bf k}}}+{V^{2}_{{\rm a}{\bf k}}\over\omega+E_{{\rm a}{\bf k}}}\right ), \label{BCSDGF}\\
\Gamma^{\dagger}({\bf k},\omega)&=&-Z_{\rm aF}{\bar{\Delta}^{({\rm a})}_{{\rm Z}{\bf k}}\over 2E_{{\rm a}{\bf k}}}\left ({1\over \omega-E_{{\rm a}{\bf k}}}-{1\over\omega
+E_{{\rm a}{\bf k}}}\right ),~~~~~\label{BCSODGF}
\end{eqnarray}
\end{subequations}
with the charge-carrier quasiparticle energy spectrum $E_{{\rm a}{\bf k}}=\sqrt{\bar{\xi}^{2}_{{\bf k}}+\mid\bar{\Delta}^{({\rm a})}_{{\rm Z}{\bf k}}\mid^{2}}$, the renormalized charge-carrier excitation spectrum $\bar{\xi}_{{\bf k}}=Z_{\rm aF}\xi_{{\bf k}}$, and the renormalized charge-carrier pair gap $\bar{\Delta}^{({\rm a})}_{{\rm Z}{\bf k}}=Z_{\rm aF} \bar{\Delta}^{({\rm a}) }_{\bf k}$, while the charge-carrier quasiparticle coherence factors,
\begin{subequations}\label{BCSCF}
\begin{eqnarray}
U^{2}_{{\rm a}{\bf k}}={1\over 2}\left (1+{\bar{\xi_{{\bf k}}}\over E_{{\rm a}{\bf k}}}\right ),\\
V^{2}_{{\rm a}{\bf k}}={1\over 2}\left (1-{\bar{\xi_{{\bf k}}}\over E_{{\rm a}{\bf k}}}\right ),
\end{eqnarray}
\end{subequations}
satisfy the constraint $U^{2}_{{\rm h}{\bf k}}+V^{2}_{{\rm h}{\bf k}}=1$ for any wave vector ${\bf k}$. In spite of the pairing mechanism driven by the kinetic energy by the exchange of spin excitations, the result in Eq. (\ref{BCSGF}) is a standard BCS expression for a charge-carrier d-wave pair state. It should be emphasized that in triangular-lattice superconductors, there is a large charge-carrier Fermi surface around the $\Gamma$ point in the Brillouin zone as well as six small hole-pockets near the $K$ points as shown in Fig. \ref{Fermi-surface}a. In particular, the nodes of the charge-carrier d-wave pair gap (\ref{d-wave-gap}) exist only on the six hole-pockets and not on the large charge-carrier Fermi surface as shown in Fig. \ref{Fermi-surface}b. However, these nodes around the six hole-pockets are far from the large charge-carrier Fermi surface, and therefore there are no gapless charge-carrier quasiparticle excitations. In this case, the effect from these charge-carrier quasiparticles around the nodes is unimportant, since everything happens at the charge-carrier Fermi surface. It is thus expected that the basic behavior of the evolution of the magnetic-field-penetration depth (then the superfluid density) with temperature in triangular-lattice superconductors is much different from that in square-lattice superconductors.

According to the full charge-carrier Green's functions in Eq. (\ref{BCSGF}) and spin Green's function in Eq. (\ref{MF-spin-Green-function}), now the charge-carrier self-energies $\Sigma^{({\rm a})}_{1}({\bf k},\omega)$ and $\Sigma^{({\rm a})}_{2}({\bf k},\omega)$ can be evaluated explicitly as,
\begin{widetext}
\begin{subequations}\label{SE1}
\begin{eqnarray}
\Sigma^{({\rm a})}_{1}({\bf k},\omega)&=&{1\over N^{2}}\sum_{{\bf pp'}n}(-1)^{n+1}\Omega^{({\rm a})}_{\bf pp'k}
\left [U^{2}_{{\rm a}{\bf p}+{\bf k}}\left ({F^{(n)}_{{\rm 1a} {\bf p p'k}}\over\omega+\omega_{n{\bf p}{\bf p}'}-E_{{\rm a}{\bf p}+{\bf k}}}+{F^{(n)}_{{\rm 2a}{\bf pp'k}} \over\omega-\omega_{n{\bf p}{\bf p}'}-E_{{\rm a}{\bf p}+{\bf k}}}\right )\right .\nonumber\\
&+&\left . V^{2}_{{\rm a}{\bf p}+{\bf k}}\left ({F^{(n)}_{{\rm 1a}{\bf pp'k}}\over\omega-\omega_{n{\bf p}{\bf p}'}+E_{{\rm a}{\bf p}+{\bf k}}}+{F^{(n)}_{{\rm 2a}{\bf pp'k}} \over \omega+\omega_{n{\bf p}{\bf p}'}+E_{{\rm a}{\bf p}+{\bf k}}}\right )\right ],~~~~~\label{PHSE}\\
\Sigma^{({\rm a})}_{2}({\bf k},\omega)&=&{1\over N^{2}}\sum_{{\bf pp'}n}(-1)^{n}\Omega^{({\rm a})}_{\bf pp'k}{\bar{\Delta}^{({\rm a})}_{{\rm Z}{\bf p}+{\bf k}}\over 2E_{{\rm a}{\bf p } +{\bf k}}}
\left [\left ({F^{(n)}_{{\rm 1a}{\bf pp'k}}\over\omega+\omega_{n{\bf p}{\bf p}'}-E_{{\rm a}{\bf p}+{\bf k}}}+{F^{(n)}_{{\rm 2a}{\bf pp'k}}\over\omega-\omega_{n{\bf p} {\bf p}'}-E_{{\rm a}{\bf p}+{\bf k}}}\right )\right .\nonumber\\
&-&\left . \left ({F^{(n)}_{{\rm 1a}{\bf pp'k}}\over\omega-\omega_{n{\bf p}{\bf p}'}+E_{{\rm a}{\bf p}+{\bf k}}}+{F^{(n)}_{{\rm 2a}{\bf pp'k}}\over\omega+\omega_{n{\bf p}{\bf p}'}+ E_{{\rm a}{\bf p}+{\bf k}}}\right )\right ], \label{PPSE}
\end{eqnarray}
\end{subequations}
respectively, with $n=1,2$, $\Omega^{({\rm a})}_{\bf pp'k}=Z_{\rm aF}(Zt\gamma_{{\bf p}+{\bf p}'+{\bf k}})^{2}B_{{\bf p}'}B_{{\bf p}+{\bf p}'}/(4\omega_{{\bf p}'}\omega_{{\bf p}+ {\bf p}'})$, $\omega_{n{\bf p}{\bf p}'}=\omega_{{\bf p}+{\bf p}'}-(-1)^{n}\omega_{\bf p'}$, and the functions,
\begin{subequations}
\begin{eqnarray}
F^{(n)}_{{\rm 1a}{\bf pp'k}}&=&n_{\rm F}(E_{{\rm a}{\bf p}+{\bf k}})\{1+n_{\rm B}(\omega_{{\bf p}'+{\bf p}})+n_{\rm B}[(-1)^{n+1}\omega_{\bf p'}]\}
+n_{\rm B}(\omega_{{\bf p}'+{\bf p}}) n_{\rm B}[(-1)^{n+1}\omega_{\bf p'}], \\
F^{(n)}_{{\rm 2a}{\bf pp'k}}&=&[1-n_{\rm F}(E_{{\rm a}{\bf p}+{\bf k}})]\{1+n_{\rm B}(\omega_{{\bf p}'+{\bf p}})+n_{\rm B}[(-1)^{n+1}\omega_{\bf p'}]\}
+n_{\rm B}(\omega_{{\bf p}'+{\bf p}}) n_{\rm B}[(-1)^{n+1} \omega_{\bf p'}],
\end{eqnarray}
\end{subequations}
where $n_{\rm B}(\omega)$ and $n_{\rm F}(\omega)$ are the boson and fermion distribution functions, respectively. In this case, the charge-carrier quasiparticle coherent weight $Z_{\rm aF}$ in Eq. (\ref{coherent-weight}) and charge-carrier pair gap parameter $\bar{\Delta}^{({\rm a})}$ in Eq. (\ref{d-wave-gap}) satisfy following two self-consistent equations,
\begin{subequations}\label{SCE1}
\begin{eqnarray}
{1\over Z_{\rm aF}}&=&1+{1\over N^{2}}\sum_{{\bf pp'}n}(-1)^{n+1}\Omega^{({\rm a})}_{{\bf pp'}{\bf k}_{0}}\left ({F^{(n)}_{{\rm 1a}{\bf pp}'{\bf k}_{0}}\over(\omega_{n{\bf p}{\bf p
}' }-E_{{\rm a}{\bf p}+{\bf k}_{0}})^{2}}
+{F^{(n)}_{{\rm 2a}{\bf pp}'{\bf k}_{0}}\over (\omega_{n{\bf p}{\bf p}'}+E_{{\rm a}{\bf p}+{\bf k}_{0}})^{2}}\right ), ~~~~~~~~\label{CCQCWSCE}\\
1&=&{6\over N^{3}}\sum_{{\bf pp'k}n}(-1)^{n}Z_{\rm aF}\Omega^{({\rm a})}_{\bf pp'k}{\Lambda^{({\rm d})*}_{\bf k}\Lambda^{({\rm d})}_{{\bf p}+{\bf k}}\over E_{{\rm a}{\bf p}+{\bf k} }}\left ({F^{(n)}_{{\rm 1a}{\bf pp'k}}\over\omega_{n{\bf p}{\bf p}'}-E_{{\rm a}{\bf p}+{\bf k}}}
-{F^{(n)}_{{\rm 2a}{\bf pp'k}}\over\omega_{n{\bf p}{\bf p}'}+E_{{\rm a}{\bf p}+{\bf k} }}\right ), ~~~~~~~~\label{CCPGPSCE}
\end{eqnarray}
\end{subequations}
\end{widetext}
respectively, with $\Lambda^{({\rm d})}_{\bf k}=d_{1{\bf k}}+id_{2{\bf k}}$. These two equations (\ref{CCQCWSCE}) and (\ref{CCPGPSCE}) must be solved simultaneously with following self-consistent equations,
\begin{subequations}\label{SCE2}
\begin{eqnarray}
\phi&=&{1\over 2N}\sum_{{\bf k}}\gamma_{{\bf k}}Z_{\rm aF}\left (1-{\bar{\xi_{{\bf k}}}\over E_{{\rm a}{\bf k}}}{\rm tanh} [{1\over 2}\beta E_{{\rm a}{\bf k}}]\right ),\\
\delta &=& {1\over 2N}\sum_{{\bf k}}Z_{\rm aF}\left (1-{\bar{\xi_{{\bf k}}}\over E_{{\rm a}{\bf k}}}{\rm tanh}[{1\over 2}\beta E_{{\rm a}{\bf k}}] \right ),~~~~\\
\chi&=&{1\over N}\sum_{{\bf k}}\gamma_{{\bf k}} {B_{{\bf k}}\over 2\omega_{{\bf k}}}{\rm coth} [{1\over 2}\beta\omega_{{\bf k}}], \\
C&=&{1\over N}\sum_{{\bf k}}\gamma^{2}_{{\bf k}} {B_{{\bf k}}\over 2\omega_{{\bf k}}}{\rm coth}[{1\over 2}\beta\omega_{{\bf k}}],\\
{1\over 2} &=&{1\over N}\sum_{{\bf k}}{B_{{\bf k}} \over 2\omega_{{\bf k}}}{\rm coth} [{1\over 2}\beta\omega_{{\bf k}}],\label{SCE2i}\\
\chi^{\rm z}&=&{1\over N}\sum_{{\bf k}}\gamma_{{\bf k}} {B_{{\rm z}{\bf k}}\over 2\omega_{{\rm z}{\bf k}}}{\rm coth}[{1\over 2}\beta\omega_{{\rm z}{\bf k}}],\\
C^{\rm z}&= &{1\over N}\sum_{{\bf k}}\gamma^{2}_{{\bf k}}{B_{{\rm z}{\bf k}}\over 2\omega_{{\rm z}{\bf k}}}{\rm coth}[{1\over 2}\beta\omega_{{\rm z}{\bf k}}],
\end{eqnarray}
\end{subequations}
then all the order parameters, the decoupling parameter $\alpha$, and the chemical potential $\mu$ are determined self-consistently without using any adjustable parameters.

\begin{figure}[h!]
\includegraphics[scale=0.3]{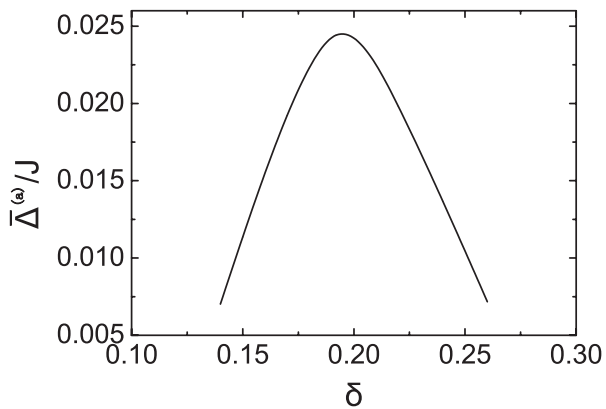}
\caption{The charge-carrier pair gap parameter as a function of doping with $T=0.001J$ for $t/J=-2.5$. \label{pair-gap-parameter-doping}}
\end{figure}

These equations in Eqs. (\ref{SCE1}) and (\ref{SCE2}) have been calculated self-consistently, and the result of the charge-carrier pair gap parameter $\bar{\Delta}^{(\rm a)}$ as a function of doping for parameter $t/J=-2.5$ with temperature $T=0.001J$ is shown in Fig. \ref{pair-gap-parameter-doping}. It is shown clearly that the charge-carrier pair gap parameter $\bar{\Delta}^{(\rm a)}$ takes a domelike shape with the underdoped and overdoped regimes on each side of the optimal doping $\delta_{\rm optimal}\approx 0.19$, where $\bar{\Delta}^{(\rm a)}$ reaches its maximum. Moreover, we have made a series of calculations for $\bar{\Delta}^{(\rm a)}$ at different temperatures, and the result shows that the charge-carrier pair gap parameter $\bar{\Delta}^{(\rm a)}$ follows qualitatively a BCS-type temperature dependence.

\begin{figure}[h!]
\includegraphics[scale=0.3]{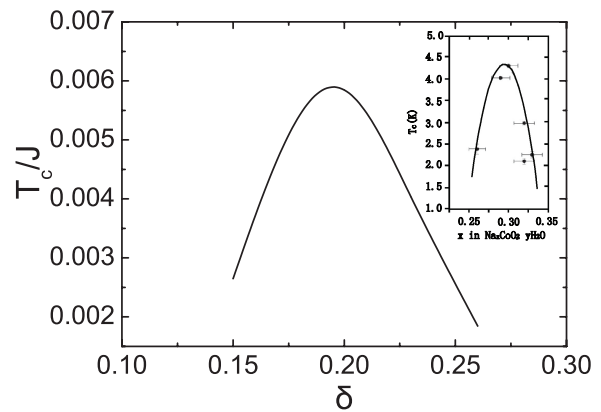}
\caption{$T_{\rm c}$ as a function of doping for $t/J=-2.5$. Inset: the corresponding experimental result of Na$_{x}$CoO$_{2}$$\cdot y$H$_{2}$O taken from Ref. \onlinecite{Schaak03} \label{Tc-doping}}
\end{figure}

$T_{\rm c}$ on the other hand can be obtained self-consistently from the self-consistent equations (\ref{SCE1}) and (\ref{SCE2}) by the condition $\bar{\Delta}^{(\rm a)}=0$, and the result of $T_{\rm c}$ as a function of doping for $t/J=-2.5$ is  plotted in Fig. \ref{Tc-doping} in comparison with the corresponding experimental result \cite{Schaak03} of Na$_{x}$CoO$_{2}$$\cdot y$H$_{2}$O (inset). Obviously, the experimental result \cite{Schaak03,Milne04,Sakurai06,Michioka06} of the doping dependence of $T_{\rm c}$ in the triangular-lattice cobaltate superconductor Na$_{x}$CoO$_{2}$$\cdot y$H$_{2}$O is qualitatively reproduced. The optimal $T_{\rm c}$ occurs in a narrow range of doping, and then decreases for both underdoped and overdoped regimes, in dramatic analogy to the phase diagram of square-lattice cuprate superconductors \cite{Tallon95}. However, in comparison with the result of $T_{\rm c}$ obtained from square-lattice cuprate superconductors, the present result also shows that the geometric frustration, accompanied by large fluctuations, suppresses $T_{\rm c}$ to low temperatures.

Within the above framework of the kinetic-energy-driven superconductivity, we now turn to discuss the doping and temperature dependence of the electromagnetic responses in triangular-lattice superconductors. For discussions of the Meissner effect in a superconductor, one usually starts from the general relation between the current and the vector potential \cite{Schrieffer64,Fukuyama69},
\begin{eqnarray}\label{current-kernel}
J_{\mu}({\bf q},\omega) &=& -\sum_{\nu} K_{\mu\nu}({\bf q},\omega)A_{\nu}({\bf q},\omega)
\end{eqnarray}
where the Greek indices label the axes of the Cartesian coordinate system, while the nonlocal kernel of the response function $K_{\mu\nu}({\bf q},\omega)$ can be expressed as $K_{\mu\nu}({\bf q},\omega)=K^{({\rm d})}_{\mu\nu}({\bf q} ,\omega)+K^{({\rm p})}_{\mu\nu}({\bf q},\omega)$, with $K^{({\rm d})}_{\mu\nu}({\bf q},\omega)$ and $K^{({\rm p})}_{\mu\nu} ({\bf q},\omega)$ that are the corresponding diamagnetic and paramagnetic parts.

In the fermion-spin representation (\ref{fermion-spin}), the vector potential ${\bf A}$ has been coupled to the electron charge which are now represented by $f_{l\uparrow}=a^{\dagger}_{l\uparrow} S^{-}_{l}$ and $f_{l\downarrow}=a^{\dagger}_{l\downarrow}S^{+}_{l}$. In this case, the electron polarization operator is expressed as
${\bf P}=e\sum_{l\sigma}{\bf R}_{l}C^{\dagger}_{l\sigma}C_{l\sigma}=e\sum_{l}{\bf R}_{l}a^{\dagger}_{l}a_{l}$, and then the corresponding electron current operator is obtained by evaluating the time derivative of this polarization operator ${\bf j}=\partial{\bf P}/\partial t=i[H,{\bf P}]/\hbar$ \cite{Feng10}. In the linear response approximation with respect to $A_{\nu}(l)$, this electron current operator is reduced as ${\bf j}={\bf j}^{({\rm d})}+{\bf j}^{({\rm p})}$, with the corresponding diamagnetic (${\rm d}$) and paramagnetic (${\rm p}$) components of the electron current operator are given by,
\begin{subequations}\label{current-1}
\begin{eqnarray}
{\bf j}^{({\rm d})}&=&-{e^{2}t\over\hbar^{2}}\sum_{l\hat{\eta}}\hat{\eta}{\bf A}(l)\cdot\hat{\eta}\left (a^{\dagger}_{l+\hat{\eta}\uparrow}a_{l\uparrow}S^{+}_{l} S^{-}_{l+\hat{\eta}} \right .\nonumber\\
&+& \left . a^{\dagger}_{l+\hat{\eta}\downarrow}a_{l\downarrow}S^{-}_{l}S^{+}_{l+\hat{\eta}}\right), \label{diamagnetic-current} \\
{\bf j}^{({\rm p})}&=&-{iet\over\hbar}\sum_{l\hat{\eta}}\hat{\eta}\left(a^{\dagger}_{l+\hat{\eta}\uparrow}a_{l\uparrow}S^{+}_{l}S^{-}_{l+\hat{\eta}} \right . \nonumber\\
&+& \left . a^{\dagger}_{l+\hat{\eta}\downarrow} a_{l\downarrow}S^{-}_{l}S^{+}_{l+\hat{\eta}}\right), \label{paramagnetic-current}
\end{eqnarray}
\end{subequations}
respectively. The diamagnetic component of the electron current operator in Eq. (\ref{diamagnetic-current}) is proportional to the vector potential, and therefore the diamagnetic part of the response kernel can be obtained directly as,
\begin{eqnarray}\label{d-kernel}
K^{({\rm d})}_{\mu\nu}({\bf q},\omega) &=& -{6e^{2}\over \hbar^{2}}\chi\phi t\delta_{\mu\nu}={1\over\lambda^{2}_{\rm L}}\delta_{\mu\nu},
\end{eqnarray}
with the London penetration depth $\lambda^{-2}_{\rm L}=-6e^{2}\chi\phi t/ \hbar^{2}$.

The paramagnetic part of the response kernel, on the other hand, is directly related to the electron current-current correlation function $P_{\mu\nu}({\bf q},\tau)=-\langle T_{\tau}\{j^{({\rm p})}_{\mu}({\bf q},\tau)j^{({\rm p})}_{\nu}(-{\bf q},0)\}\rangle$, and can be expressed as $K^{({\rm p})}_{\mu\nu}({\bf q},\omega)=P_{\mu\nu}({\bf q},\omega)$. In the fermion-spin approach, the paramagnetic component of the electron current operator in Eq. (\ref{paramagnetic-current}) can be decoupled as,
\begin{eqnarray}\label{paramagnetic-current-1}
{\bf j}^{({\rm p})}&=&-{ie\chi t\over\hbar}\sum_{l\hat{\eta}\sigma}\hat{\eta}a^{\dagger}_{l+\hat{\eta}\sigma}a_{l\sigma}\nonumber\\
&-&{ie\phi t\over\hbar}\sum_{l\hat{\eta}}\hat{\eta} (S^{+}_{l} S^{-}_{l+\hat{\eta}}+S^{-}_{l}S^{+}_{l+\hat{\eta}}).
\end{eqnarray}
As in the case of square-lattice superconductors \cite{Feng10}, the second term in the right-hand side refer to the contribution from the electron spin, and can be shown that $\sum\limits_{l\hat{\eta}}\hat{\eta}(S^{+}_{l}S^{-}_{l+\hat{\eta}}+S^{-}_{l}S^{+}_{l+\hat{\eta}})\equiv 0$, i.e., there is no direct contribution for the electron current-current correlation function $P_{\mu\nu}({\bf q},\tau)$ from the electron spin, and then the majority contribution for $P_{\mu\nu}({\bf q},\tau)$ comes from the electron charge, however the strong interplay between charge carriers and spins has been considered through the spin's order parameters entering in the charge-carrier part of the contribution to the current-current correlation $P_{\mu\nu}({\bf q},\tau)$.

\begin{figure}[h!]
\includegraphics[scale=0.3]{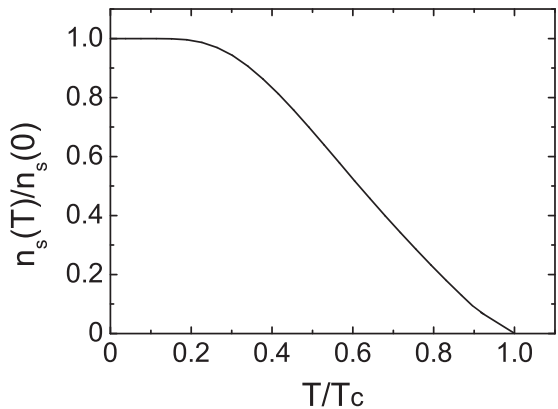}
\caption{The effective superfluid density as a function of temperature at $\delta=0.15$ for $t/J=-2.5$. \label{effective-superfluid-density}}
\end{figure}

The density operator is summed over the position of all particles, i.e, $\rho(l)=-e[1/N]\sum_{l\sigma}C^{\dagger}_{l\sigma}C_{l\sigma}=-e[1/(2N)]\sum_{l\sigma}a^{\dagger}_{l\sigma} a_{l\sigma}$, and then its Fourier transform can be expressed as $\rho({\bf q})=-e/(2N)\sum_{{\bf k}\sigma} a^{\dagger}_{{\bf k} \sigma}a_{{\bf k}+ {\bf q}\sigma}$. For a convenience in the following discussions, the paramagnetic component of the electron current operator in Eq. (\ref{paramagnetic-current}) and density operator can be rewritten into the four-current operator in the Nambu representation in terms of the charge-carrier Nambu operators $\Psi^{\dagger}_{k}= (a^{\dagger}_{k\uparrow}, a_{-k\downarrow})$ and $\Psi_{k+q}=(a_{k+q\uparrow},a^{\dagger}_{-k-q\downarrow})^{\rm T}$ as,
\begin{eqnarray}
j^{({\rm p})}_{\mu}({\bf q})={1\over N}\sum_{\bf k}\Psi^{\dagger}_{\bf k}\gamma_{\mu}({\bf k},{\bf k}+{\bf q})\Psi_{{\bf k}+{\bf q}}, \label{paramagnetic-current-2}
\end{eqnarray}
with the bare current vertex,
\begin{widetext}
\arraycolsep=0.15em\begin{eqnarray}\label{current-vertex}
\gamma_{\mu}({\bf k},{\bf k}+{\bf q}) &=&
\begin{cases}\medmuskip=0mu\thinmuskip=0mu
-{2e\chi t\over\hbar}e^{{1\over 4}iq_{x}}\{\cos{1\over 4}q_{x}\sin(k_{x}+{1\over 2}q_{x})+\cos{\sqrt{3}\over 4}q_{y}\sin({1\over 2}k_{x}+{1\over 4}q_{x})\cos({\sqrt{3}\over 2}k_{y}+{\sqrt{3}\over 4}q_{y})\\
\medmuskip=0mu\thinmuskip=0mu +i[\sin{1\over 4}q_{x}\sin(k_{x}+{1\over 2}q_{x})+\sin{\sqrt{3}\over 4}q_{y}\cos({1\over 2}k_{x}+{1\over 4}q_{x})\sin({\sqrt{3}\over 2}k_{y}+{\sqrt{3}\over 4}q_{y})]\} \mspace{126mu} \text{for \thickmuskip=0mu $\mu=x$} \\
\medmuskip=0mu\thinmuskip=0mu -\sqrt{3}{2e\chi t\over\hbar}e^{i{\sqrt{3}\over 4}q_{y}}[\cos{1\over 4}q_{x}\cos({1\over 2}k_{x}+{1\over 4}q_{x})\sin({\sqrt{3}\over 2}k_{y}+{\sqrt{3}\over 4}q_{y})\\
\medmuskip=0mu\thinmuskip=0mu +i\sin{1\over 4}q_{x}\sin({1\over 2}k_{x}+{1\over 4}q_{x})\cos({\sqrt{3}\over 2}k_{y}+{\sqrt{3}\over 4}q_{y})] \mspace{310mu} \text{for \thickmuskip=0mu $\mu=y$} \\
-{1\over 2}e\tau_{3} \mspace{578mu} \text{for \thickmuskip=0mu $\mu=0$}
\end{cases}
\end{eqnarray}
In this case, the current-current correlation function is obtained as,
\begin{eqnarray}
P_{\mu\nu}({\bf q},i\omega_{n})&=&{1\over N}\sum\limits_{\bf k}\gamma_{\mu}({\bf k}+{\bf q},{\bf k})\gamma^{*}_{\nu}({\bf k}+{\bf q},{\bf k})
{1\over\beta}\sum\limits_{i\nu_{m}}{\rm Tr}[\tilde{g}({\bf k}+{\bf q},i\omega_{n}+i\nu_{m})\tilde{g}({\bf k},i\nu_{m})],~~~~~~~~\label{correlation-function-1}
\end{eqnarray}
where the full charge-carrier Green's function $\tilde{g}({\bf k},\omega)$ in the Nambu representation can be expressed in terms of the full charge-carrier Green's function (\ref{BCSGF}) as,
\begin{eqnarray}
\tilde{g}({\bf k},\omega)=Z_{\rm aF}{\omega\tau_{0}+\bar{\xi}_{{\bf k}}\tau_{3}-\bar{\Delta}^{({\rm a})}_{\rm Z}(d_{1{\bf k}}\tau_{1}+d_{2{\bf k}}\tau_{2})\over\omega^{2}-
E^{2}_{{\rm a}{\bf k}}}. \label{NPBCSHG}
\end{eqnarray}
Substituting this charge-carrier Green's function (\ref{NPBCSHG}) into Eq. (\ref{correlation-function-1}), the paramagnetic part of the response kernel in the static limit ($\omega\sim 0$) is evaluated as,
\begin{eqnarray}\label{p-kernel}
K^{({\rm p})}_{\mu\nu}({\bf q},0)&=&{1\over N}\sum_{\bf k}\gamma_{\mu}({\bf k}+{\bf q},{\bf k})\gamma^{*}_{\nu}({\bf k}+{\bf q},{\bf k})[L^{({\rm a})}_{1}({\bf k},{\bf q})
+L^{({\rm a})}_{2}({\bf k},{\bf q}) ]
= K^{({\rm p})}_{\mu\mu}({\bf q},0)\delta_{\mu\nu},~~~~
\end{eqnarray}
with the functions $L^{({\rm a})}_{1}({\bf k},{\bf q})$ and $L^{({\rm a})}_{2}({\bf k},{\bf q})$ are given by,
\begin{subequations}
\begin{eqnarray}\label{lkq}
L^{({\rm a})}_{1}({\bf k},{\bf q})&=& Z^{2}_{\rm aF}\left(1+{\bar{\xi}_{\bf k}\bar{\xi}_{{\bf k}+{\bf q}}+{1\over 2}\bar{\Delta}^{(\rm a)}_{{\rm Z}{\bf k}}\bar{\Delta}^{(\rm a)*}_{ {\rm Z}{\bf k}+{\bf q}}+{1\over 2}\bar{\Delta}^{(\rm a)*}_{{\rm Z}{\bf k}}\bar{\Delta}^{(\rm a)}_{{\rm Z}{\bf k}+{\bf q}}\over E_{{\rm a}{\bf k}}E_{{\rm a}{\bf k}+{\bf q}}}\right)  {n_{\rm F}(E_{{\rm a}{\bf k}})- n_{\rm F}(E_{{\rm a}{\bf k}+{\bf q}})\over E_{{\rm a}{\bf k}}-E_{{\rm a}{\bf k}+{\bf q}}},\\
L^{({\rm a})}_{2}({\bf k},{\bf q}) &=& Z^{2}_{\rm aF}\left(1-{\bar{\xi}_{\bf k}\bar{\xi}_{{\bf k}+{\bf q}}+{1\over 2}\bar{\Delta}^{(\rm a)}_{{\rm Z}{\bf k}}\bar{\Delta}^{(\rm a)* }_{{\rm Z}{\bf k}+{\bf q}}+{1\over 2}\bar{\Delta}^{(\rm a)*}_{{\rm Z}{\bf k}}\bar{\Delta}^{(\rm a)}_{{\rm Z}{\bf k}+{\bf q}}\over E_{{\rm a}{\bf k}}E_{{\rm a}{\bf k}+{\bf q}}} \right) {n_{\rm F}(E_{{\rm a}{\bf k}})+ n_{\rm F}(E_{{\rm a}{\bf k}+{\bf q}})-1\over E_{{\rm a}{\bf k}}+E_{{\rm a}{\bf k}+{\bf q}}},
\end{eqnarray}
\end{subequations}
respectively. In this case, the kernel of the response function in Eq. (\ref{current-kernel}) is now obtained from Eqs. (\ref{d-kernel}) and (\ref{p-kernel}) as,
\begin{eqnarray}\label{total-kernel}
K_{\mu\nu}({\bf q},0) &=& \left[{1\over\lambda^{2}_{\rm L}}+K^{({\rm p})}_{\mu\mu}({\bf q},0)\right]\delta_{\mu\nu}.
\end{eqnarray}
In the long-wavelength limit, i.e., $|{\bf q}|\to 0$, the function $L^{({\rm a})}_{2}({\bf k},{\bf q}\to 0)$ vanishes, then the paramagnetic part of the response kernel in Eq. (\ref{p-kernel}) is reduced as,
\begin{eqnarray}\label{p-kernel-1}
K^{({\rm p})}_{\rm yy}({\bf q}\to 0,0)&=&Z^{2}_{\rm aF}{24e^{2}\over\hbar^{2}}{1\over N}\sum_{\bf k}\chi^{2}t^{2}\cos^{2}({1\over2}k_{x})\sin^{2}({\sqrt{3}\over 2}k_{y})
\lim_{{\bf q} \to 0}{n_{\rm F}(E_{{\rm a}{\bf k}})-n_{\rm F}(E_{{\rm a}{\bf k}+{\bf q}})\over E_{{\rm a}{\bf k}}-E_{{\rm a}{\bf k}+{\bf q}}}. ~~~~~~~
\end{eqnarray}
However, at zero temperature ($T=0$), $K^{({\rm p})}_{\rm yy}({\bf q}\to 0,0)|_{T=0}=0$, and then the long-wavelength electromagnetic response is determined by diamagnetic part of the response kernel $K^{({\rm d})}_{\rm yy}$ only. On the other hand, at $T=T_{\rm c}$, the charge-carrier gap parameter $\bar{\Delta}^{(\rm a)}|_{T=T_{\rm c}}=0$, and in this case, the paramagnetic part of the response kernel in the long-wavelength limit can be evaluated as,
\begin{eqnarray}
K^{({\rm p})}_{\rm yy}({\bf q}\to 0,0)&=&Z^{2}_{\rm aF}{24e^{2}\over\hbar^{2}}{1\over N}\sum_{\bf k}\chi^{2}t^{2}\cos^{2}({1\over2}k_{x})\sin^{2}({\sqrt{3}\over 2}k_{y})\lim_{{\bf q} \to 0}{n_{\rm F}(\bar{\xi}_{\bf k})-n_{\rm F}(\bar{\xi}_{{\bf k}+{\bf q}})\over \bar{\xi}_{\bf k}-\bar{\xi}_{{\bf k}+{\bf q}}}
=-{1\over\lambda^{2}_{\rm L}},
\end{eqnarray}
which exactly cancels the diamagnetic part of the response kernel in Eq. (\ref{d-kernel}), and then the Meissner effect in triangular-lattice superconductors disappears for all temperatures $T\geq T_{\rm c}$. These results also reflect that the Meissner effect is strongly temperature dependent. To show this point clearly, we introduce an effective superfluid density $n_{\rm s}(T)$ at temperature $T$, which is defined in terms of the paramagnetic part of the response kernel as,
\begin{eqnarray}
K^{({\rm p})}_{\mu\nu}({\bf q}\to 0,0) &=& -{1\over \lambda^{2}_{\rm L}}\left [1-{n_{\rm s}(T)\over n_{\rm s}(0)}\right ]\delta_{\mu\nu},
\end{eqnarray}
where the ratio $n_{\rm s}(T)/n_{\rm s}(0)$ of the effective superfluid densities at temperature $T$ and zero-temperature is obtained directly from the paramagnetic part of the response kernel (\ref{p-kernel-1}) as,
\begin{eqnarray}\label{superfluid-density}
{n_{\rm s}(T)\over n_{\rm s}(0)}&=&1-\lambda^{2}_{\rm L}Z^{2}_{\rm aF}{24e^{2}\over\hbar^{2}}{1\over N}\sum_{\bf k}[\chi t\cos({1\over 2}k_{x})\sin({\sqrt{3}\over 2}k_{y})]^{2}{\beta e^{\beta E_{{\rm a}{\bf k}}}\over (e^{\beta E_{{\rm a}{\bf k}}}+1)^{2}}.
\end{eqnarray}
\end{widetext}
In this case, the kernel of the response function in Eq. (\ref{total-kernel}) can be expressed explicitly in terms of the effective superfluid density as,
\begin{eqnarray}
K_{\mu\nu}({\bf q}\to 0,0) &=& {1\over\lambda^{2}_{\rm L}}{n_{\rm s}(T)\over n_{\rm s}(0)}\delta_{\mu\nu}. \label{kernel2}
\end{eqnarray}
In Fig. \ref{effective-superfluid-density}, we plot this effective superfluid density $n_{s}(T)/n_{s}(0)$ as a function of temperature at $\delta=0.15$ for $t/J=-2.5$, where $n_{s}(T)/n_{s}(0)$ decreases with increasing temperatures, and vanishes at $T_{\rm c}$, then all the charge-carriers are in the normal fluid for temperatures $T \ge T_{\rm c}$. To sum up, within the kinetic-energy driven SC mechanism, we find that: (a) the Meissner effect in triangular-lattice superconductors is obtained for all temperatures $T\leq T_{\rm c}$ throughout the SC dome; (b) the electromagnetic response kernel goes to the London form in the long-wavelength limit [see, e.g., Eq. (\ref{kernel2})]; (c) although the electromagnetic response kernel (\ref{total-kernel}) is not manifestly gauge invariant within the bare current vertex (\ref{current-vertex}), however, we can keep the gauge invariance within the dressed current vertex as it has been done in the case for square-lattice superconductors \cite{Krzyzosiak10}; (d) in spite of the pairing mechanism driven by the kinetic energy by the exchange of spin excitations, the kinetic-energy-driven SC-state in triangular-lattice superconductors still is conventional BCS-like with the d-wave symmetry without gap nodes at the charge-carrier Fermi surface, which leads to a fact that the superfluid density therefore follows essentially a BCS-type temperature dependence as in the case of conventional superconductors \cite{Schrieffer64}.

\section{Doping dependence of electromagnetic response} \label{meissner-effect}

In the above discussions, it is shown that once the response kernel $K_{\mu\nu}$ is known, the effect of a weak electromagnetic field on a superconductor can be quantitatively characterized by experimentally measurable quantities such as the magnetic-field-penetration depth and superfluid density. However, the result of the effective superfluid density in Eq. (\ref{superfluid-density}) obtained from the response kernel in Eq. (\ref{kernel2}) can not be used for a direct comparison with the corresponding experimental data of triangular-lattice superconductors because the kernel function derived within the linear response theory describes the response of an infinite system \cite{Feng10,Krzyzosiak10}, whereas in the actual problem of the penetration of the field and the system has a surface, i.e., it occupies a half-space $x>0$. In this case, we need to impose boundary conditions for charge carriers, which can be done within the simplest specular reflection model \cite{Abrikosov88,Tinkham96} with a two-dimensional geometry of the SC plane. Following our previous discussions of the electromagnetic response in square-lattice superconductors \cite{Feng10,Krzyzosiak10}, the local magnetic field profile of triangular-lattice superconductors can be evaluated explicitly as,
\begin{eqnarray}\label{local-magnetic}
h_{\rm z}(x) &=& {B\over \pi}\int^{\infty}_{-\infty}{\rm d}q_{x}{q_{x}\sin(q_{x}x)\over \mu_{0}K_{\rm yy}(q_{x},0,0)+q^{2}_{x}},
\end{eqnarray}
and then the magnetic-field-penetration depth is obtained from this local-magnetic-field profile as,
\begin{eqnarray}\label{penetration-depth}
\lambda(T) &=& {1\over B}\int^{\infty}_{0}h_{\rm z}(x){\rm d}x\nonumber\\
&=&{2\over \pi}\int^{\infty}_{0}{{\rm d}q_{x}\over \mu_{0}K_{\rm yy}(q_{x},0,0)+q^{2}_{x}}.
\end{eqnarray}
For a convenience in the following discussions, we introduce a characteristic length scale $a_{0}=\sqrt{\hbar^{2}a/ \mu_{0}e^{2}J}$. Using the lattice parameter $a\approx 0.282$ nm for Na$_{x}$CoO$_{2}$$\cdot$ yH$_{2}$O, this characteristic length is obtained as $a_{0}\approx 83.9$ nm.

\begin{figure}[h!]
\includegraphics[scale=0.3]{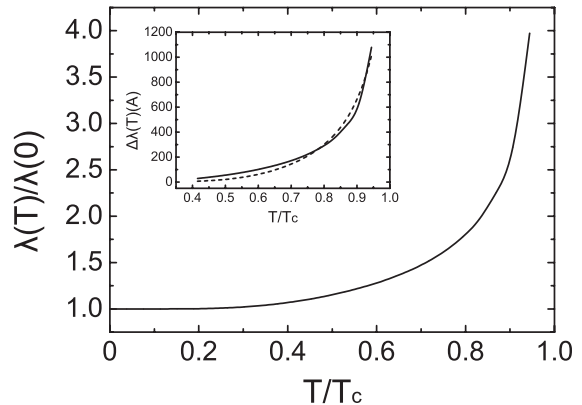}
\caption{The magnetic-field-penetration as a function of temperature at $\delta=0.15$ for $t/J=-2.5$. Inset: the magnetic-field-penetration difference $\Delta\lambda(T)=\lambda(T)-\lambda(0)$ (solid line) as a function of temperature at $\delta=0.15$ for $t/J=-2.5$, while the dashed line is obtained from a numerical fit $\Delta\lambda(T)=A\exp[-B\bar{\Delta}^{({\rm a})}(T) /T]$ with $A\sim 2789.62$ and $B\sim 0.59$. \label{penetration-depth-doping}}
\end{figure}

We are now ready to discuss the electromagnetic response in triangular-lattice superconductors. At zero temperature, the obtained magnetic-field-penetration depths from Eq. (\ref{penetration-depth}) are $\lambda(0)\approx 362.96$ nm, $\lambda(0)\approx 316.38$ nm, and $\lambda(0)\approx 294.36$ nm for $\delta=0.15$, $\delta=0.17$, and $\delta=0.19$, respectively. However, at $T=T_{\rm c}$, since the kernel of the response function $K_{\mu\nu}({\bf q}\to 0,0)|_{T=T_{c}}=0$, the magnetic-field-penetration depth is found as $\lambda(T_{c})=\infty$, i.e., the external magnetic field can penetration through all the main body of the system for $T\geq T_{\rm c}$, and then the Meissner effect does not exist in the normal-state. On the other hand, $\lambda(T)$ is sensitive to low-lying excitations. To show this point clearly, $\lambda(T)/\lambda(0)$ as a function of temperature at $\delta=0.15$ for $t/J=-2.5$ is plotted in Fig. \ref{penetration-depth-doping}. It is seen that below temperatures $T<0.25T_{\rm c}$, $\lambda(T)$ is practically independent temperature, which is a reflection of the absence of the d-wave gap nodes at the large charge-carrier Fermi surface. However, above temperatures $T>0.25T_{\rm c}$, $\lambda(T)$ increases rapidly with increasing temperature. In particular, we have fitted our present theoretical result of the magnetic-field-penetration depth difference $\Delta\lambda(T)=\lambda(T)-\lambda(0)$, and the fitted result is shown in inset of Fig. \ref{penetration-depth-doping}. We thus find that $\Delta\lambda(T)$ vary exponentially as a function of temperature ($\Delta\lambda(T)= A\exp[-B\bar{\Delta}^{({\rm a})}(T)/T]$ with $A\sim 2789.62$ and $B\sim 0.59$), which is expected result in the case without the d-wave gap nodes at the large charge-carrier Fermi surface.

An external magnetic field acts on the SC-state of triangular-lattice superconductors as a perturbation. In the linear response form (\ref{current-kernel}), the nonlocal relation between the supercurrent and the vector potential in the coordinate space holds due to the finite size of charge-carrier pairs. In particular, the size of charge-carrier pairs is of the order of the coherence length $\zeta({\bf k})=\hbar v_{\rm F}/(\pi\bar{\Delta}^{(\rm a)}_{\bf k})$, where $v_{\rm F}=\hbar^{-1}\partial\xi_{\bf k}/\partial {\bf k}|_{k_{\rm F}}$ is the charge-carrier velocity at the large charge-carrier Fermi surface, which shows that the size of charge-carrier pairs is momentum dependent. In general case, although the external magnetic field decays on the scale of the magnetic-field-penetration length $\lambda(T)$, any nonlocal contributions to measurable quantities are of the order of $\kappa^{-2}$, where the Ginzburg--Landau parameter $\kappa$ is the ratio of the magnetic-field-penetration depth $\lambda$ and the coherence length $\zeta$. However, for the charge-carrier d-wave pair gap (\ref{d-wave-gap}), there is no gap nodes at the large charge-carrier Fermi surface. In this case, the momentum dependent coherence length $\zeta({\bf k})$ can be replaced approximately with the isotropic one $\zeta_{0}=\hbar v_{\rm F}/(\pi\bar{\Delta}^{(\rm a)})$, and then the condition for the local limit is satisfied. As a consequence, triangular-lattice superconductors are type-II superconductors due to the existence of the anisotropic energy gap over the large charge-carrier Fermi surface, where nonlocal effects are negligible, and then the electrodynamics is purely local and the magnetic field decays exponentially over a length of the order of a few hundreds nm. In this local limit, the pure d-wave pairing state in the kinetic-energy-driven SC mechanism gives a temperature dependence of the magnetic-field-penetration depth as $\Delta\lambda(T)\propto\exp[-\bar{\Delta}^{({\rm a})}(T)/T]$. This is much different from the case in square-lattice cuprate superconductors, where the characteristic feature of the d-wave charge-carrier pair gap is the existence of the four nodes at the charge-carrier Fermi surface, and then the quasiparticle excitations are gapless and affect particularly the physical properties at the extremely low temperatures. These gapless quasiparticle excitations in square-lattice cuprate superconductors lead to a divergence of the coherence length $\zeta({\bf k})$ around the gap nodes, and then the behavior of the temperature dependence of the magnetic-field-penetration depth depends sensitively on the quasiparticle scattering. At the extremely low temperatures, the quasiparticles selectively locate around the gap nodal region, and then the major contribution to measurable quantities comes from these quasiparticles. In this case, the Ginzburg--Landau ratio $\kappa({\bf k})$ around the gap nodal region is no longer large enough for the system to belong to the class of type-II superconductors, and the condition of the local limit is not satisfied \cite{Feng10,Feng15}, which leads to the system in the extreme nonlocal limit, and therefore the nonlinear behavior in the temperature dependence of the magnetic-field-penetration depth is observed experimentally \cite{Suter04}. On the other hand, with increasing temperatures, the quasiparticles around the gap nodal region become excited out of the condensate, and then the nonlocal effect fades away, which leads to that the magnetic-field-penetration depth crossovers to the linear temperature dependence \cite{Suter04}. However, the present result of the temperature dependence of the magnetic-field-penetration depth in triangular-lattice superconductors is very similar to the case in conventional superconductors \cite{Schrieffer64}, where the characteristic feature is the existence of the isotropic energy gap at the Fermi surface, and then the temperature dependence of the magnetic-field-penetration depth exhibits an exponential behavior.

\begin{figure}[h!]
\includegraphics[scale=0.3]{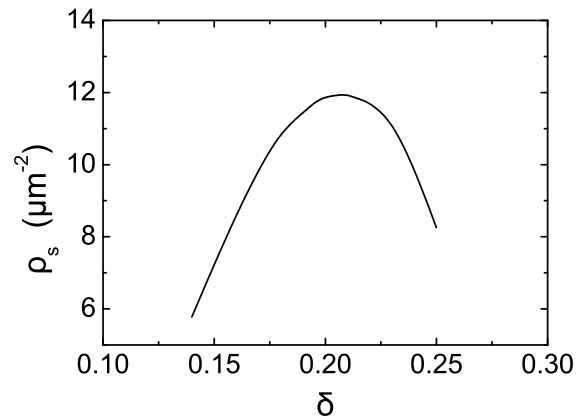}
\caption{The superfluid density as a function of doping with $T=0.001J$ for $t/J=-2.5$. \label{superfluid-density-doping}}
\end{figure}

Now we turn to discuss the doping dependence of the superfluid density $\rho_{\rm s}(T)$, which is a measure of the phase stiffness, and is defined in terms of the magnetic-field-penetration depth $\lambda(T)$ as $\rho_{\rm s}(T)\equiv 1/\lambda^{2}(T)$. In this case, we have performed firstly a calculation for the doping dependence of $\rho_{\rm s}$ in triangular-lattice superconductors for all levels of doping throughout the SC dome, and the result is plotted in Fig. \ref{superfluid-density-doping}. In analogy to the dome-like shape of the doping dependence of $T_{\rm c}$ shown in Fig. \ref{Tc-doping}, $\rho_{\rm s}$ also displays a dome-like shape of the doping dependence, i.e., it increases with increasing doping in the lower doped regime, and reaches a maximum (a peak) around the {\it critical doping} $\delta_{\rm critical}\approx 0.21$, then decreases in the higher doped regime. Moreover, $\rho_{\rm s}$ of triangular-lattice superconductors in the underdoped regime vanishes more or less linearly with decrease of the charge-carrier doping concentration $\delta$. In square-lattice cuprate superconductors, one of the most unconventional natures is that $T_{\rm c}$ scales with $\rho_{\rm s}$ following the so-called Uemura relation as $T_{\rm c}\propto const\times \rho_{\rm s}$ in the underdoped regime \cite{Uemura91}. It is interesting to know if triangular-lattice superconductors also obey this relation. In this case, we have fitted the relation between $T_{\rm c}$ and $\rho_{\rm s}$ in the underdoped regime, and the result shows that triangular-lattice superconductors satisfy the similar Uemura relation in the underdoped regime \cite{Kanigel04,Uemural04}. Incorporating the result obtained from square-lattice cuprate superconductors \cite{Feng10}, it thus implies that the Uemura relation may be a universal relation in strongly correlated superconductors in despite of whether the gap nodes at the charge-carrier Fermi surface exist or not.

The essential physics of the dome-like shape of the doping dependence of $\rho_{\rm s}$ in triangular-lattice superconductors is the same as in the case of square-lattice superconductors \cite{Feng10,Feng15}, and also can be attributed to the dome-like shape of the doping dependence of $\bar{\Delta}^{(\rm a)}$. This follows a fact that $\rho_{\rm s} (T)$ in triangular-lattice superconductors is intriguingly related to the current-current correlation function, and therefore the variation of the superfluid density with doping is coupled to the doping dependence of the charge-carrier pair gap parameter $\bar{\Delta}^{(\rm a)}$. In particular, the charge-carrier pair gap parameter $\bar{\Delta}^{(\rm a)}$ measures the strength of the binding of two charge carriers into a charge-carrier pair. On the other hand, the superfluid density $\rho_{\rm s}$ is a measurement of the phase stiffness, and is proportional to the squared amplitude of the charge-carrier pair macroscopic wave functions. In this case, both $\rho_{\rm s}$ and $\bar{\Delta}^{(\rm a)}$ describe the different aspects of the same charge-carrier quasiparticles, and then the dome-like shape of the doping dependence of $\rho_{\rm s}$ in Fig. \ref{superfluid-density-doping} is a natural consequence of the dome-like shape of the doping dependence of $\bar{\Delta}^{(\rm a)}$ shown in Fig. \ref{pair-gap-parameter-doping}.

\begin{figure}[h!]
\includegraphics[scale=0.3]{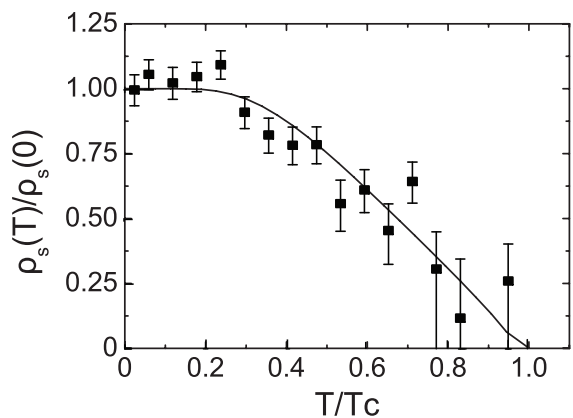}
\caption{The superfluid density as a function of temperature at $\delta=0.15$ for $t/J=-2.5$. The experimental result of Na$_{x}$CoO$_{2}\cdot y$H$_{2}$O (solid squares) taken from Ref. \onlinecite{Uemural04}. \label{superfluid-density-temp}}
\end{figure}

The superfluid density shown in Fig. \ref{superfluid-density-doping} also is strongly temperature dependence. When the temperature $T=T_{\rm c}$, the kernel of the response function $K_{\mu\nu}({\bf q}\to 0,0)|_{T=T_{\rm c}}=0$, and then $\lambda(T_{\rm c})=\infty$ as mentioned above, which leads to $\rho_{\rm s}(T_{\rm c})=0$, and is consistent with the result of the effective superfluid density obtained from Eq. (\ref{superfluid-density}). For a better understanding of the basic behavior of $\rho_{\rm s}(T)$ as a function of temperature, we have made a series of calculations for $\rho_{\rm s}(T)$ at different temperatures, and the result of $\rho_{\rm s}(T)$ as a function of temperature at $\delta=0.15$ for  $-t/J=2.5$ is plotted in Fig. \ref{superfluid-density-temp} in comparison with the corresponding experimental result \cite{Uemural04} of Na$_{x}$CoO$_{2}\cdot y$H$_{2}$O (solid square). Our present calculations thus qualitatively reproduce the overall evolution of the superfluid density with temperature in Na$_{x}$CoO$_{2}\cdot y$H$_{2}$O \cite{Uemural04}. In corresponding to the result of the temperature dependence of the magnetic-field-penetration depth shown in Fig. \ref{penetration-depth-doping}, $\rho_{s}(T)$ is also independence of temperature below temperatures $T<0.25T_{\rm c}$, and then decreases dramatically with increasing temperature for temperatures $T>0.25T_{\rm c}$, eventually vanishing together with superconductivity at $T_{\rm c}$. The calculation based on the kinetic-energy-driven SC mechanism with the d-wave charge-carrier pair gap (\ref{d-wave-gap}) thus gives a good agreement with the observed superfluid density data of Na$_{x}$CoO$_{2}\cdot y$H$_{2}$O.

\section{Conclusions}\label{conclusions}

Within the framework of the kinetic-energy driven superconductivity, we have performed a calculation of the doping and temperature dependence of the Meissner effect in triangular-lattice superconductors for all temperatures $T\le T_{\rm c}$ throughout the SC dome. Our results indicate that the magnetic-field-penetration depth shows an exponential temperature dependence due to the absence of the d-wave gap nodes at the large charge carrier Fermi surface. In particular, the experimental result of the temperature dependence of the superfluid density in cobaltate superconductors can be qualitatively described in terms of the d-wave pairing state. However, as a natural consequence of the dome-like shape of the doping dependence of the charge-carrier pair gap parameter and $T_{\rm c}$, the superfluid density increases with increasing doping in the lower doped regime, and reaches a highest value (a peak) around the critical doping, then decreases in the higher doped regime.

\acknowledgments

This work was supported by the funds from the Ministry of Science and Technology of China under Grant Nos. 2011CB921700 and 2012CB821403, and the National Natural Science Foundation of China under Grant Nos. 11274044 and 11447144. JQ is supported by the Fundamental Research Funds for the Central Universities under Grant No. FRF-TP-14-074A2, and the Beijing Higher Education Young Elite Teacher Project under Grant No. 0389.

\end{document}